\documentclass[onecolumn,aps,superscriptaddress,showpacs,nofootinbib,floatfix]{revtex4}
\usepackage{listings}
\usepackage[utf8]{inputenc}
\usepackage{hyperref}
\usepackage{amsmath}
\usepackage{amssymb}
\usepackage{amsfonts}
\usepackage{booktabs}
\usepackage{tabularx}
\usepackage{graphicx}
\usepackage{color}
\usepackage{float}
\usepackage{supertabular}
\usepackage{calc}
\usepackage{array}
\usepackage{seqsplit}
\usepackage{changepage}
\usepackage{multirow,makecell}
\usepackage[inline]{enumitem}
\usepackage[T2A,T1]{fontenc}
\graphicspath{{fig/}}
\definecolor{theblue}{RGB}{0,50,230}

\usepackage{hyperref}
\hypersetup{
  colorlinks=true,
  linkcolor=theblue,
  citecolor=theblue,
  urlcolor=theblue
}

\newcommand {\avg}[1]{\ensuremath{\langle\kern-1.0pt\langle#1\rangle\kern-1.0pt\rangle}}

\newlength\cmsFigWidth

\newcommand{\PreserveBackslash}[1]{\let\temp=\\#1\let\\=\temp}
\usepackage[normalem]{ulem}  % \sout{old text} for strikeout

\renewcommand\sout{\bgroup \color{red} \ULdepth=-.5ex \ULset}
\begin{document}

%%%%%%%%%%%%%%%%%%%%% Title %%%%%%%%%%%%%%%%%%%%%%

\title{The Novel Scaling of Tsallis Parameters from the Transverse Momentum Spectra of Charged Particles in Heavy-Ion Collisions}

%%%%%%%%%%%%%%%%%%%% Authors %%%%%%%%%%%%%%%%%%%%%
\author{J.Q. Tao}
\affiliation{School of Physics and Information Technology, Shaanxi Normal University, Xi'an 710119, China}

\author{W.H. Wu}
\affiliation{School of Physics and Information Technology, Shaanxi Normal University, Xi'an 710119, China}

\author{M. Wang}
\affiliation{School of Physics and Information Technology, Shaanxi Normal University, Xi'an 710119, China}

\author{H. Zheng}
\affiliation{School of Physics and Information Technology, Shaanxi Normal University, Xi'an 710119, China}

\author{W.C. Zhang}
\affiliation{School of Physics and Information Technology, Shaanxi Normal University, Xi'an 710119, China}

\author{L.L. Zhu}
\affiliation{Department of Physics, Sichuan University, Chengdu 610064, China}

\author{A. Bonasera}
\affiliation{Cyclotron Institute, Texas A\&M University, College Station, Texas 77843, USA}
\affiliation{Laboratori Nazionali del Sud, INFN, via Santa Sofia, 62, 95123 Catania, Italy}

%%%%%%%%%%%%%%%%%%%% Abstract %%%%%%%%%%%%%%%%%%%%%

\begin{abstract}
The transverse momentum $(p_T)$ spectra of charged particles measured in Au + Au collisions from the beam energy scan (BES) program, Cu + Cu collisions at $\sqrt{s_{NN}}=62.4$, 200 GeV at the RHIC and Pb + Pb, Xe + Xe collisions at the LHC are investigated in the framework of Tsallis thermodynamics. The theory can describe the experimental data well for all the collision systems, energies and centralities investigated. The collision energy and centrality dependence of the Tsallis distribution parameters, i.e., the temperature $T$ and the nonextensive parameter $q$, for the \mbox{A + A} collisions are also studied and discussed. A novel scaling between the temperature divided by the natural logarithm of collision energy $(T/\ln\sqrt{s})$ and the nonextensive parameter $q$ is presented.
\end{abstract}

\maketitle

\section{Introduction}
In experiments, the transverse momentum ($p_T$) spectra of particles produced in the high energy heavy-ion collisions can be measured and provide information about the collision system. They are used to constrain and improve the transport models designed to study the heavy-ion collisions, as well as to gain deep insights into the collision processes and to extract the kinetic freeze-out information, {such as the kinetic freeze-out temperature and volume}, of the collision system. Abundant data of the transverse momentum spectra of particles have been measured by different experimental collaborations at the RHIC and LHC in the past two decades, ranging from p+p collisions to Pb + Pb collisions at different energies and centralities \cite{dataAdamczyk2018,dataAdler2002,dataBack2004,dataAlver2006,dataAcharya2018,dataAcharya2019,ptAad2016,ptAbe1988,ptAbelev2009,ptAdams2003,ptAdam2015,ptAdam2016PLB753,ptAdam2016PLB760,ptAdamczyk2017,ptAdams2004,ptArnison1982,ptBack2003,ptBack2005,ptKhachatryan2010}. 

Some well-established models \cite{modelLin2005, modelOstapchenko2011, modelWang1991, modelSa2012, modelNara2018, modelBozek2012,modelIvanov2006, modelFries2003PRL,modelFries2003PRC, RMHwa20032, RMHwa2003, RMZhu2021,RMHwa2018, tan1, tan2}, such as AMPT, QGSJET \uppercase\expandafter{\romannumeral2}, HIJING, PACIAE, JAM, Hydrodynamic model, and the recombination model, are used to reproduce the transverse momentum spectra of particles and understand the dynamical evolution of the collision system and the relevant physics. Many other theoretical models have also been proposed to describe the transverse momentum spectra of particles, such as the blast-wave (BW) model \cite{thmodelSchnedermann1993}, the Tsallis blast-wave model \cite{thmodelChe2021}, the two-component model \cite{thmodelLiu2014}, and so on.

Recently, the Tsallis distribution has attracted lots of attention due to its successful applications to the particle transverse momentum spectra and the pseudorapidity distribution of charged particles produced in high energy heavy-ion collisions \cite{Azmi2015,Azmi2020,Marques2015,Gao2017,Zheng2015PRD,Tao2021,Patra2021,Sarawat2018,Cleymans2012JPG,Wang2019,Cleymans2012EPJA, Parvan2017,Cleymans2017,Khandai2013, Waqas2021EPJP, Biro:2020kve}. In the present study, we apply the Tsallis distribution to systematically analyze the transverse momentum spectra of charged particles produced in Au + Au collisions from the beam energy scan (BES) program, i.e., collision energies ranging from $\sqrt{s_{NN}}=7.7$ to 200 GeV, Cu + Cu collisions at $\sqrt{s_{NN}}=62.4$ and 200 GeV, Pb + Pb collisions at $\sqrt{s_{NN}}=2.76$ and 5.02 TeV and Xe + Xe collisions at $\sqrt{s_{NN}}=5.44$ TeV. We also study the collision energy and centrality dependence of Tsallis distribution parameters, i.e., the temperature $T$ and nonextensive parameter $q$, and search for some novel phenomena.

The remainder of this paper is organized as follows. In Section \ref{sec2}, we briefly introduce the Tsallis distribution for the transverse momentum spectra of charged particles. In \mbox{Section \ref{sec3}}, we show the results along with the experimental data generated in different heavy-ion collision systems at the RHIC and LHC. Furthermore, a novel scaling between the ratio $T/\ln\sqrt{s}$ and $q$ is found for all the collision systems investigated. In Section \ref{sec4}, a brief conclusion is drawn.

\section{Formula of the Transverse Momentum Distribution}\label{sec2}
As a generalization of the Boltzmann–Gibbs distribution in classical thermodynamics, the Tsallis distribution was proposed several decades ago \cite{Tsallis1988}. The transverse momentum distribution of particles derived from the Tsallis distribution can be written as \cite{Azmi2020,Marques2015,Gao2017,Tao2021}
\begin{eqnarray}
	\frac{d^{2} N}{2 \pi p_{T} d p_{T} d y}=g V \frac{m_{T} \cosh y}{(2 \pi)^{3}}\left[1+(q-1) \frac{m_{T} \cosh y-\mu}{T}\right]^{-\frac{q}{q-1}},\label{1}
\end{eqnarray}
where $g$ is particle state degeneracy, $m_{T}=\sqrt{m^{2}_{0}+p^{2}_{T}}$ is the transverse mass and $m_{0}$ is the particle rest mass, $y$ is the rapidity and $\mu$ is the chemical potential. $V$ is the volume, $T$ is the temperature and $q$ is {the entropic factor which measures the nonadditivity of the \mbox{entropy \cite{Tsallis1988, tfl}}}. When $q=1$, Equation (\ref{1}) recovers the Boltzmann distribution. For the collision energies explored in this work, the multiplicities of charged particles and their antiparticles are approximately equal, therefore we take $\mu=0$. The experimental data are taken in the mid-rapidity, $y\approx0$; thus, Equation (\ref{1}) can be rewritten as
\begin{eqnarray}
	\frac{d^{2} N}{2 \pi p_{T} d p_{T} d y}=g V \frac{m_{T}}{(2 \pi)^{3}}\left[1+(q-1) \frac{m_{T}}{T}\right]^{-\frac{q}{q-1}}. \label{2}
\end{eqnarray}

In order to be consistent with the experimental data, we convert the rapidity to pseudorapidity in Equation (\ref{2}). The relation between rapidity and pseudorapidity {in the mid-rapidity} is
\begin{eqnarray}
	\frac{d y}{d \eta}=\sqrt{1-\frac{m_{0}^{2}}{m_{T}^{2} \cosh ^{2} y}}.\label{3}
\end{eqnarray}

As it is well known, most of the charged particles produced in high energy heavy-ion collisions are $\pi^{+}(\pi^{-})$, $K^{+}(K^{-})$ and $p(\overline{p})$. Therefore, we can sum over the Tsallis distributions of these particles at the mid-rapidity to fit the transverse momentum distribution of charged particles produced in heavy-ion collisions. Then, we obtain \cite{Azmi2020}
\begin{eqnarray}
	\frac{d^{2} N_{ch}}{2 \pi p_{T} d p_{T} d\eta}=\sum_{i}g_{i} V \frac{p_T}{(2 \pi)^{3}}\left[1+(q-1) \frac{m_{T, i}}{T}\right]^{-\frac{q}{q-1}}, \label{4}
\end{eqnarray}
where $i=\pi, K, p$. The degeneracy factor $g_{i}$ are $g_{\pi} = g_{K} = 2$ and $g_{p} = 4$ to take into account the spin and the contribution from antiparticles. We noted multiplicity differences between protons and anti-protons at low collision energies but {these} are only a small portion of the total charged particles. Equation (\ref{4}) will be used to extract the parameters $T$ and $q$ by fitting the experimental transverse momentum distribution of charged particles.

\section{Results}\label{sec3}
\begin{figure}[H]
\centering
\includegraphics[width=14.5cm]{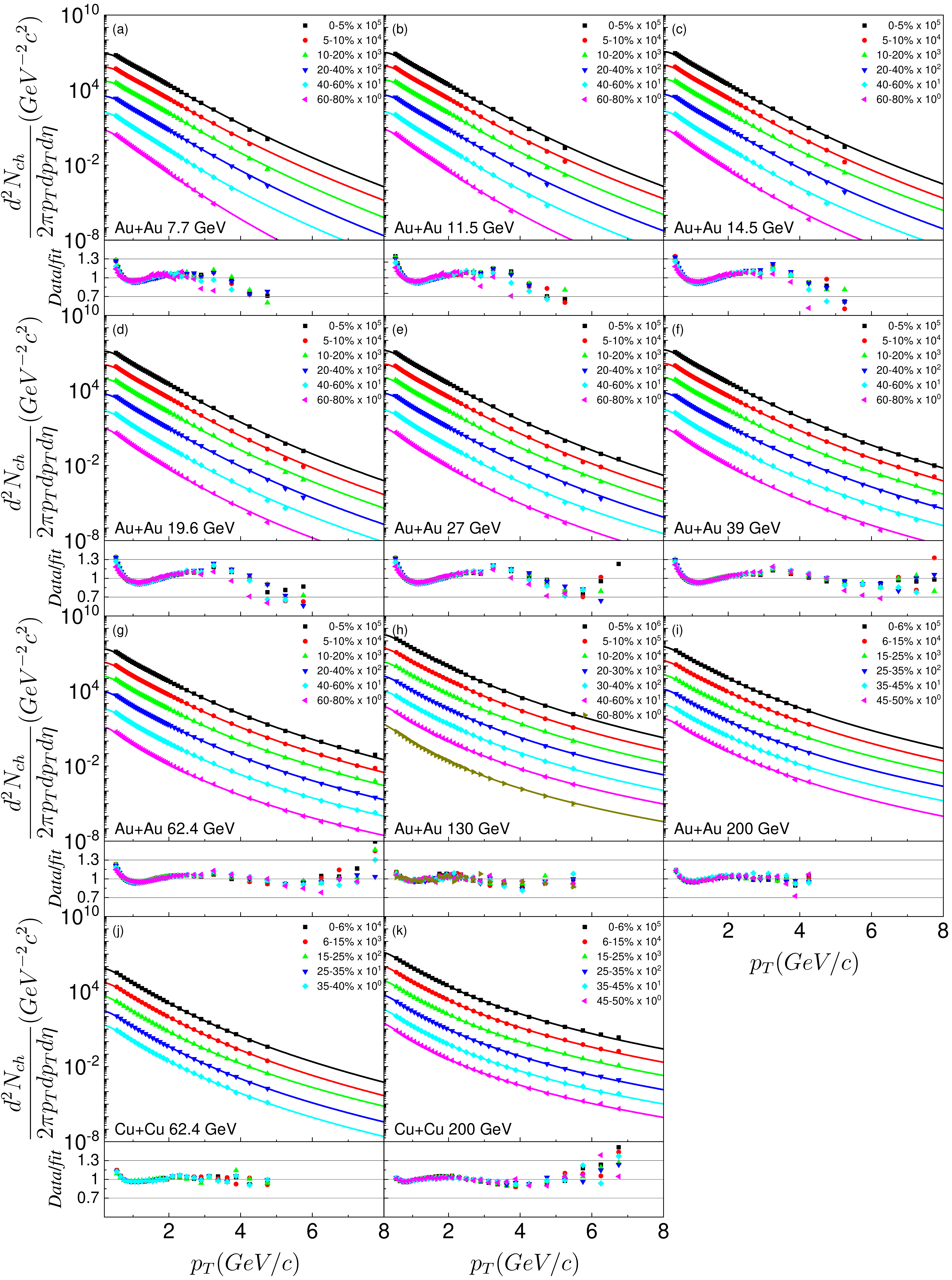}
\caption{(Color online) The transverse momentum distribution of charged particles produced in Au + Au collisions for energies ranging from $\sqrt{s_{NN}}=7.7$ to 200 GeV and Cu + Cu collisions at $\sqrt{s_{NN}}=62.4$ and 200 GeV at the RHIC. Scale factors are applied for better visibility. The curves are the fit results from Equation (\ref{4}). The experimental data are taken from Refs. \cite{dataAdamczyk2018,dataAdler2002,dataBack2004,dataAlver2006}. The corresponding ratios of data/fit are also shown.}\label{RHIC_pt}
\end{figure}

\begin{figure}[H]
\centering
		\includegraphics[width=14.5cm]{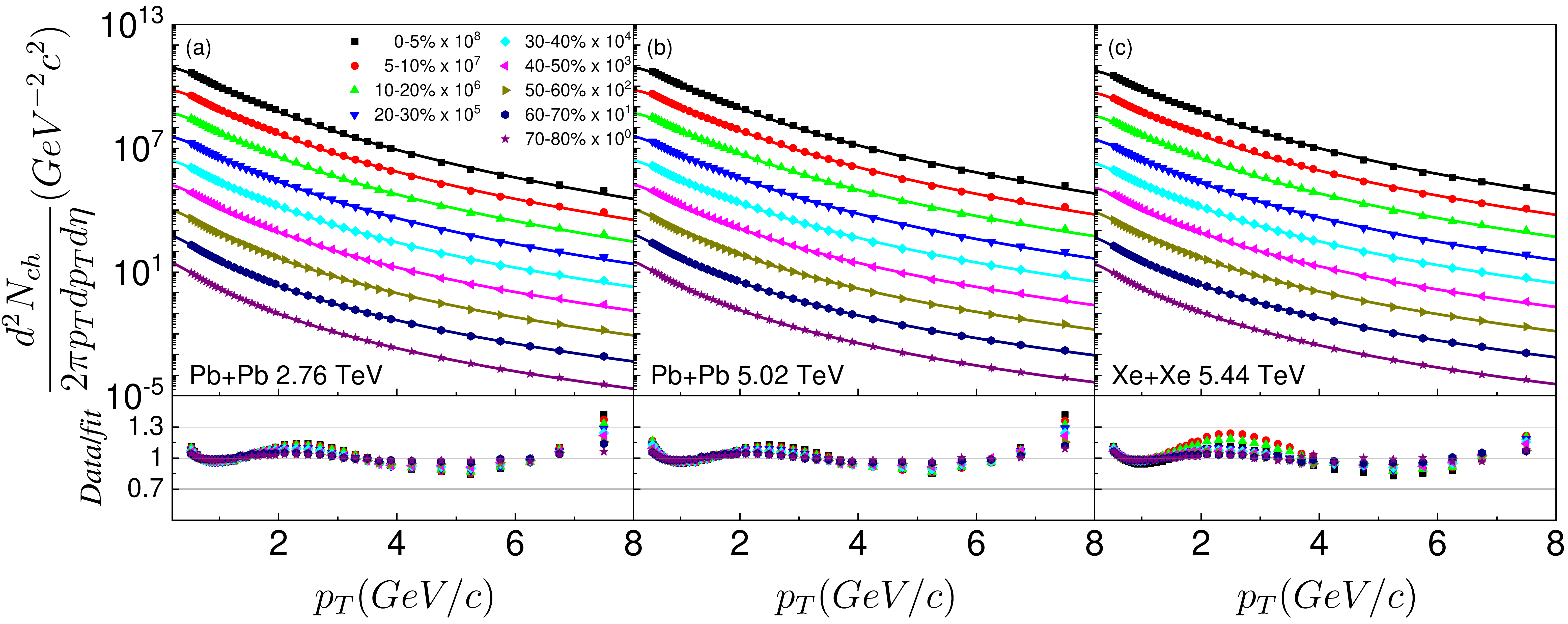}
	\caption{(Color online) Same as Figure \ref{RHIC_pt} but for Pb + Pb collisions at $\sqrt{s_{NN}}=2.76$ and 5.02 TeV and Xe + Xe collisions at $\sqrt{s_{NN}}=5.44$ TeV at LHC. The experimental data are taken from Refs. \cite{dataAcharya2018,dataAcharya2019}.}\label{LHC_pt}
\end{figure}

In Figures \ref{RHIC_pt} and \ref{LHC_pt} {are displayed} the results of {the} Tsallis distribution, \mbox{Equation (\ref{4})}, to fit the transverse momentum distribution of charged particles at different centralities in \mbox{Au + Au} collisions with collision energy (BES program) from $\sqrt{s_{NN}}=7.7$ to \mbox{200 GeV,} \mbox{Cu + Cu} collisions at $\sqrt{s_{NN}}=62.4$ and 200 GeV at the RHIC, Pb + Pb collisions at $\sqrt{s_{NN}}=2.76$ and 5.02 TeV as well as Xe + Xe collisions at $\sqrt{s_{NN}}=5.44$ TeV at LHC. {The corresponding $\chi^2$/NDF for the fits are listed in the Appendix \ref{app1}.} The Tsallis distribution can well reproduce the experimental data of all the collision systems and centralities investigated. Noting that we have applied the cut $p_T<8$ GeV/c to the transverse momentum spectra of charged particles in Pb + Pb and Xe + Xe collisions at LHC \cite{Azmi2020}. To show the agreement between the data and the Tsallis distribution in linear scale, the ratios of data/fit are shown in the bottom panels of Figures \ref{RHIC_pt} and \ref{LHC_pt}. The discrepancies are within 30\% for the low collision energies and it reduces to 15\% for the high collision energies. The centrality dependence of the fitting quality is not obvious at the RHIC. However, the Tsallis distribution can fit the transverse momentum spectrum better at the peripheral collisions than at the central collisions for the collision systems at LHC. This can be attributed to the medium effects {for} central collisions, because it is empirically known that the Tsallis distribution can fit {well} the transverse 
momentum spectra produced in p + p collisions. The peripheral collisions are more like the p + p collisions than {the} central collisions \cite{Azmi2015,Zheng2015PRD}.

\begin{figure}[H]
\centering
	\includegraphics[width=8.5cm]{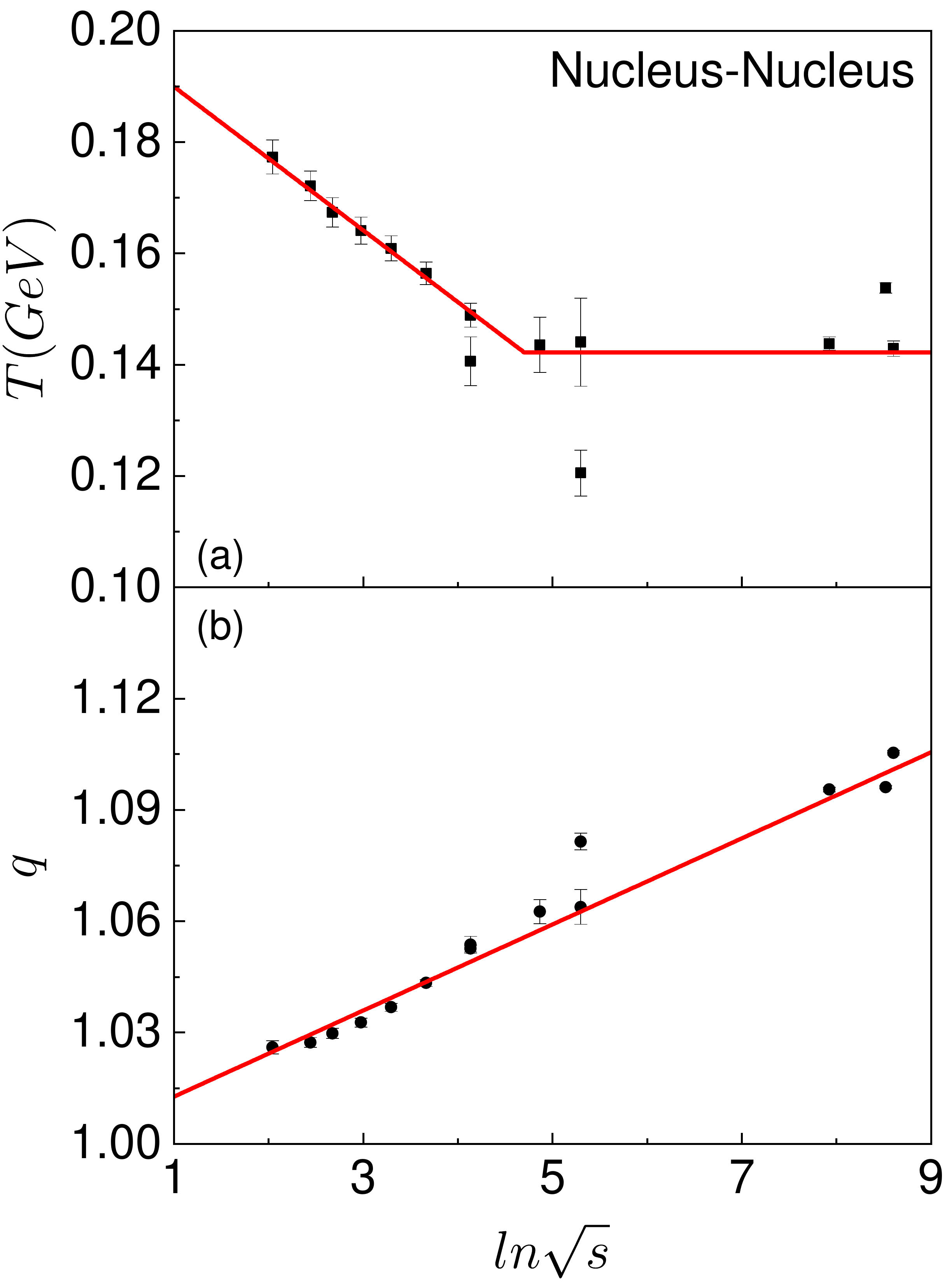}
	\caption{(Color online) The collision energy dependence of the temperature $T$ and nonextensive parameter $q$ for the collision systems in Figures \ref{RHIC_pt} and \ref{LHC_pt} at the most central collisions. See text for \mbox{the lines.}}\label{parameters_vs_lnsqrts}
\end{figure}

We analyze the collision energy and centrality dependence of the Tsallis distribution parameters, i.e., the temperature $T$ and the nonextensive parameter $q$. In Figure \ref{parameters_vs_lnsqrts}, the results of $T$ and $q$ versus collision energy from the RHIC to LHC for the most central collisions are shown. {We use $\sqrt{s}$ to denote $\sqrt{s_{NN}}$ in units of GeV, i.e., 
\begin{equation}
\sqrt{s}=\sqrt{s_{NN}}/1 \text{GeV},
\end{equation};
thus, it is a dimensionless variable that is suitable used in an expression like $\ln\sqrt{s}$.} For the temperature $T$, it is observed that a linear decrease from $\sqrt{s_{NN}}=7.7$ GeV to a certain collision energy and then it is approximately constant for the higher collision energies, {which is the asymptotic value connected to the Hagedorn temperature, i.e., the pion mass}. The lines are drawn to guide the eyes. {Unlike} the temperature $T$, the parameter $q$ shows a linear monotonic increasing dependence on the collision energy in the whole energy region investigated. A linear fit gives $q=0.0116\ln\sqrt{s}+1.00116$ shown in the Figure \ref{parameters_vs_lnsqrts}b. This indicates that the higher the collision energy is, the less the collision system reaches thermal equilibrium during the evolution and the temperature fluctuation is larger \cite{tfl}. The parameters from the other centralities showing similar behaviors have been observed.

\begin{figure}[H]
\centering
	\includegraphics[width=14.5cm]{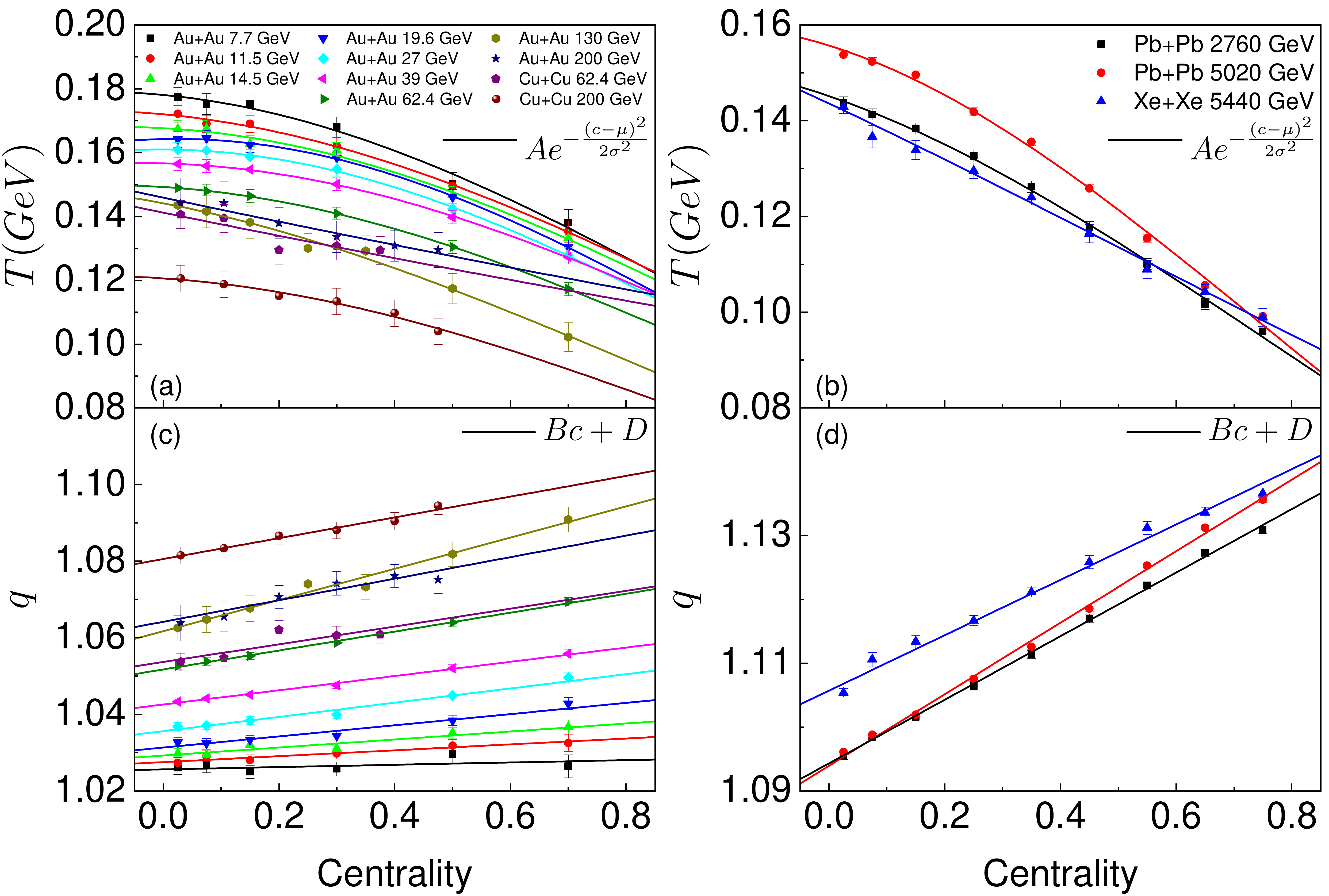}
	\caption{(Color online) The centrality (0 represents the most central collisions) dependence of the $T$ and $q$ in Au + Au collisions at $\sqrt{s_{NN}}=7.7-200$ GeV, Cu + Cu collisions at $\sqrt{s_{NN}}=62.4$, 200 GeV, \mbox{Pb + Pb} collisions $\sqrt{s_{NN}}=2.76$, 5.02 TeV and Xe + Xe collisions at $\sqrt{s_{NN}}=5.44$ TeV. The curves in (\textbf{a},\textbf{b}) are the parabolic fits and the lines in (\textbf{c},\textbf{d}) are the linear fits.}\label{parameters_vs_centrality}
\end{figure}

In Figure \ref{parameters_vs_centrality}, the results of the centrality dependence of $T$ and $q$ for A + A collisions at different collision energies are presented. A nice parabolic fit can be performed for temperature $T$ and centrality, see the solid lines in Figure \ref{parameters_vs_centrality}a,b. The corresponding parameters are listed in Table \ref{t1}. {Unlike} the temperature $T$, the dependence between $q$ and {the} centrality is linear as shown by the linear fits in Figure \ref{parameters_vs_centrality}c,d. The fit parameters are listed in Table \ref{t1} as well. The observation is that the temperature $T$ decreases from central to peripheral collisions {for} all collision systems and energies as expected. This trend is universal and is consistent with the average momentum dependence of centrality \cite{avpt}. The system can survive for a longer time in central collisions than in peripheral collisions since more nucleons are affected and more particles are produced in the collisions. Therefore more collisions can occur resulting in the higher temperatures for central collisions. The nonextensive parameter $q$ increases from central to peripheral collisions, having the opposite trend respect to the temperature $T$ versus centrality. Similarly to the discussion above, it indicates that the peripheral collisions are further away from thermal equilibrium and have larger temperature fluctuations than the central collisions for the same collision system and energy.

\begin{figure}[H]
\centering
	\includegraphics[width=12cm]{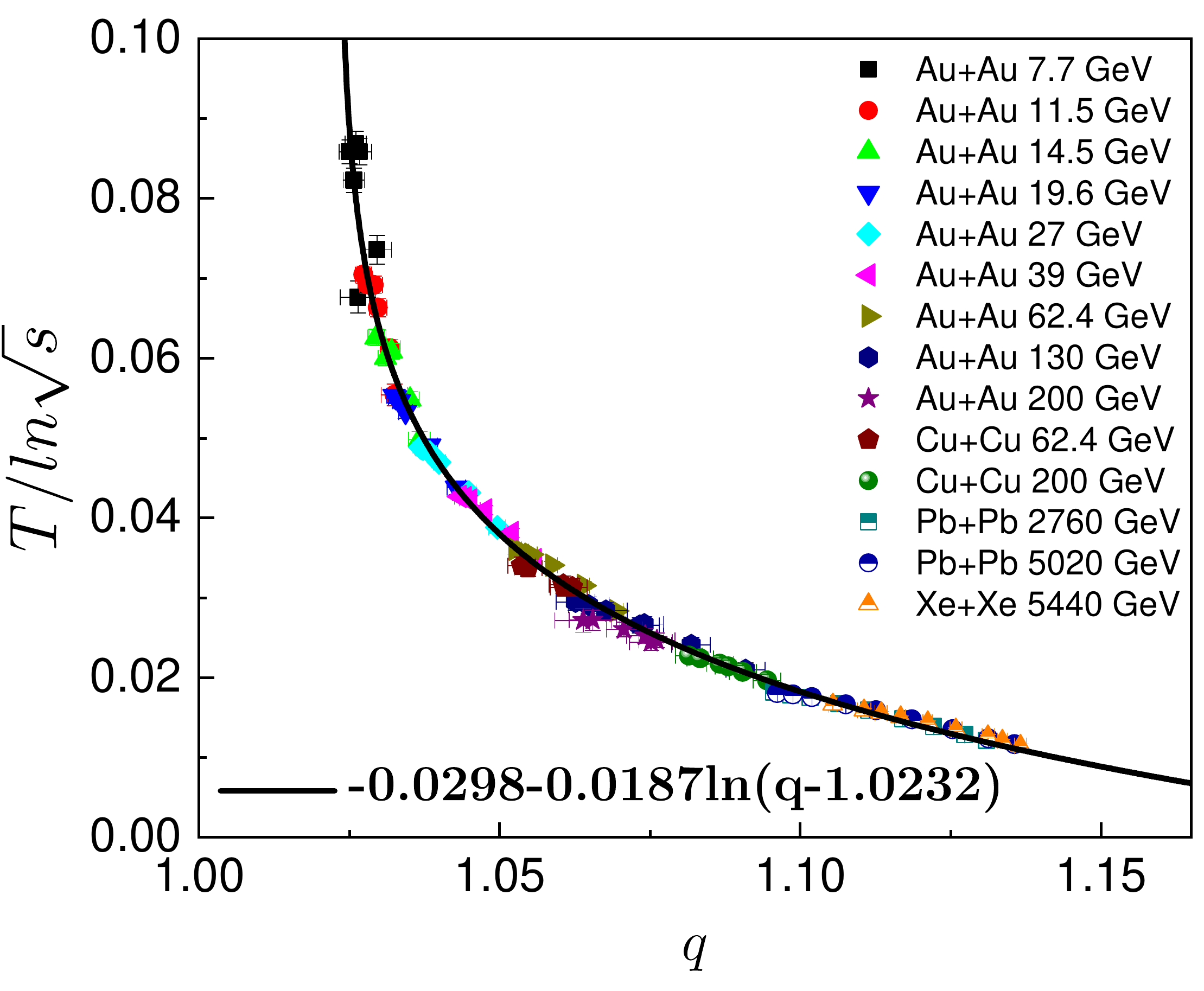}
	\caption{(Color online) Nonextensive parameter $q$ dependence of the temperature divided by the natural logarithm of collision energy $T/ln\sqrt{s}$ in A + A collisions with different centrality. The curve is the fit results and the fit function is indicated in the legend.}\label{new_scaling}
\end{figure}

In Figure \ref{new_scaling}, the results of the novel scaling discovered between the temperature divided by the natural logarithm of collision energy $(T/\ln\sqrt{s})$ and nonextensive parameter $q$ for all the A + A collision systems and centralities investigated is shown. It is clear that all the data points are scaled into one curve. We are able to fit it with the function indicated in the legend of Figure \ref{new_scaling}. This {observed strong scaling} indicates that the parameters of Tsallis distribution obtained from the charged particle transverse momentum spectra are not independent of each other but are anticorrelated. {It also suggests that further fundamental characteristics of the nonextensive statistics are yet to be studied.} The emergence of this scaling maybe attributed to hydrodynamical scaling \cite{bcprl} but further investigations \mbox{are needed.}

\renewcommand\arraystretch{1.5}
\begin{table}[H]
	\centering
	\caption{The fit parameters of $T$ and $q$ as function of centrality $c$.}
	\begin{tabular}{>{\centering}p{2.5cm}>{\centering}p{2.5cm}>{\centering}p{7.0cm}p{3cm}<{\centering}}
		\hline
		System&$\sqrt{s_{NN}}$ (GeV)&$T$ (GeV)&$q$\\
		\hline
		\multirowcell{9}{Au+Au}&7.7  &$-0.0514(c+0.2338)^2+0.1813$&$0.0030c+1.0256$\\
		&11.5 &$-0.0456(c+0.2336)^2+0.1747$&$0.0077c+1.0275$\\
		&14.5 &$-0.0504(c+0.1484)^2+0.1690$&$0.0104c+1.0293$\\
		&19.6 &$-0.0669(c+0.0177)^2+0.1646$&$0.0145c+1.0313$\\
		&27   &$-0.0599(c+0.0517)^2+0.1614$&$0.0186c+1.0356$\\
		&39   &$-0.0478(c+0.0963)^2+0.1573$&$0.0186c+1.0425$\\
		&62.4 &$-0.0432(c+0.1842)^2+0.1509$&$0.0247c+1.0517$\\
		&130  &$-0.0296(c+0.6607)^2+0.1573$&$0.0407c+1.0617$\\
		&200  &$-0.0039(c+4.4090)^2+0.2213$&$0.0281c+1.0642$\\
		\hline
		\multirowcell{2}{Cu+Cu}&62.4 &$-0.0038(c+4.4077)^2+0.2159$&$0.0232c+1.0537$\\
		&200  &$-0.0421(c+0.1577)^2+0.1217$&$0.0271c+1.0806$\\
		\hline
		\multirowcell{2}{Pb+Pb}&2760 &$-0.0239(c+1.0463)^2+0.1719$&$0.0498c+1.0943$\\
		&5020 &$-0.0368(c+0.7023)^2+0.1748$&$0.0560c+1.0940$\\
		\hline
		\multirowcell{1}{Xe+Xe}&5440 &$-0.0052(c+5.4808)^2+0.2991$&$0.0434c+1.1057$\\
		\hline
	\end{tabular}\label{t1}
\end{table}

\section{Conclusions}\label{sec4}

In this paper, the transverse momentum spectra of charged particles produced at different collision systems, energies and centralities at the RHIC and LHC have been studied by using Tsallis distribution. The results show that Tsallis distribution can fit well all the charged particle spectra in a wide range of $p_T$ investigated. In the centrality and collision energy dependence analysis, it is found that the temperature $T$ linearly decreases with $\ln\sqrt{s}$ to a certain collision energy and becomes almost a constant for a selected centrality while the nonextensive parameter $q$ linearly increases with $\ln\sqrt{s}$. For a fixed collision energy, the temperature decreases from central to peripheral collisions and can be fitted with a parabolic function and the parameter $q$ linearly increases from central to peripheral collisions indicating that the peripheral collisions are further away from the thermal equilibrium and have larger temperature fluctuations referring to central collisions. Furthermore, we have found a novel scaling between the temperature divided by the natural logarithm of the collision energy $T/\ln\sqrt{s}$ and the nonextensive parameter $q$. More works related to nonextensive statistics may be stimulated by this finding.

\section*{Acknowledgements}
This research was funded in part by the National Natural Science Foundation of China (Grant Nos. 11905120 and 11947416) and by the United States Department of Energy under Grant \# DE-FG03-93ER40773 and the NNSA Grant No. DENA0003841 (CENTAUR).

\newpage
\appendix
\section[\appendixname \thesection]{The $\chi^2/NDF$ of the Fitting Results in Figures %MDPI: Figure citations are not allowed in the heading; please move them to the main text or delete them.
 \ref{RHIC_pt} and \ref{LHC_pt}}\label{app1}

\renewcommand\arraystretch{1.5}
\begin{center}
	\centering
	\topcaption{The $\chi^2$ of the fitting result in Fig.1.}
	\begin{supertabular}{>{\centering}p{5cm}>{\centering}p{5cm}p{5cm}<{\centering}}
		\hline
		System&Centrality&$\chi^2/NDF$\\
		\hline
		\multirowcell{6}{Au + Au 7.7 GeV}&  0-5\%&0.789\\
		& 5-10\%&0.762\\
		&10-20\%&0.837\\
		&20-40\%&0.684\\
		&40-60\%&0.385\\
		&60-80\%&0.263\\
		\hline
		\multirowcell{6}{Au + Au 11.5 GeV}&  0-5\%&1.223\\
		& 5-10\%&1.018\\
		&10-20\%&1.162\\
		&20-40\%&1.045\\
		&40-60\%&0.720\\
		&60-80\%&0.463\\
		\hline
		\multirowcell{6}{Au + Au 14.5 GeV}&  0-5\%&1.183\\
		& 5-10\%&1.366\\
		&10-20\%&1.077\\
		&20-40\%&1.391\\
		&40-60\%&1.023\\
		&60-80\%&0.671\\
		\hline
		\multirowcell{6}{Au + Au 19.6 GeV}&  0-5\%&1.260\\
		& 5-10\%&1.444\\
		&10-20\%&1.424\\
		&20-40\%&1.618\\
		&40-60\%&1.380\\
		&60-80\%&0.775\\
		\hline
		\multirowcell{6}{Au + Au 27 GeV}&  0-5\%&1.256\\
		& 5-10\%&1.392\\
		&10-20\%&1.368\\
		&20-40\%&1.464\\
		&40-60\%&1.207\\
		&60-80\%&0.820\\
		\hline
		\multirowcell{6}{Au + Au 39 GeV}&  0-5\%&0.929\\
		& 5-10\%&0.998\\
		&10-20\%&1.030\\
		&20-40\%&1.072\\
		&40-60\%&1.032\\
		&60-80\%&1.039\\
		\hline
		\multirowcell{6}{Au + Au 62.4 GeV}&  0-5\%&0.698\\
		& 5-10\%&0.665\\
		&10-20\%&0.711\\
		&20-40\%&0.640\\
		&40-60\%&0.632\\
		&60-80\%&0.574\\
		\hline
		\multirowcell{7}{Au + Au 130 GeV}&  0-5\%&0.302\\
		& 5-10\%&0.321\\
		&10-20\%&0.231\\
		&20-30\%&0.182\\
		&30-40\%&0.246\\
		&40-60\%&0.165\\
		&60-80\%&0.174\\
		\hline
		\multirowcell{6}{Au + Au 200 GeV}&  0-6\%&0.142\\
		& 6-15\%&0.188\\
		&15-25\%&0.255\\
		&25-35\%&0.247\\
		&35-45\%&0.345\\
		&45-50\%&0.467\\
		\hline
		\multirowcell{5}{Cu + Cu 62.4 GeV}&  0-6\%&0.199\\
		& 6-15\%&0.236\\
		&15-25\%&0.178\\
		&25-35\%&0.213\\
		&35-40\%&0.480\\
		\hline
		\multirowcell{6}{Cu + Cu 200 GeV}&  0-6\%&0.531\\
		& 6-15\%&0.377\\
		&15-25\%&0.290\\
		&25-35\%&0.189\\
		&35-45\%&0.266\\
		&45-50\%&0.264\\
		\hline
	\end{supertabular}\label{t2}
\end{center}

\newpage 

\begin{center}
	\centering
	\topcaption{The $\chi^2$ of the fitting result in Fig.2.}
	\begin{supertabular}{>{\centering}p{5cm}>{\centering}p{5cm}p{5cm}<{\centering}}
		\hline
		System&Centrality&$\chi^2/NDF$\\
		\hline
		\multirowcell{9}{Pb + Pb 2.76 TeV}&  0-5\%&12.669\\
		& 5-10\%&10.693\\
		&10-20\%&9.489\\
		&20-30\%&7.168\\
		&30-40\%&5.862\\
		&40-50\%&4.095\\
		&50-60\%&2.990\\
		&60-70\%&1.794\\
		&70-80\%&0.957\\
		\hline
		\multirowcell{9}{Pb + Pb 5.02 TeV}&  0-5\%&22.199\\
		& 5-10\%&20.963\\
		&10-20\%&21.037\\
		&20-30\%&18.503\\
		&30-40\%&16.016\\
		&40-50\%&12.836\\
		&50-60\%&9.049\\
		&60-70\%&6.126\\
		&70-80\%&3.204\\
		\hline
		\multirowcell{9}{Xe + Xe 5.44 TeV}&  0-5\%&8.279\\
		& 5-10\%&2.777\\
		&10-20\%&3.238\\
		&20-30\%&3.425\\
		&30-40\%&2.796\\
		&40-50\%&1.279\\
		&50-60\%&0.709\\
		&60-70\%&0.995\\
		&70-80\%&0.481\\
		\hline
	\end{supertabular}\label{t3}
\end{center}

\end{document}